\documentclass[a4paper,aps,prd,twocolumn,groupedaddress,amssymb,eqsecnum,showpacs,nofootinbib]{revtex4}
\usepackage{graphicx}
\usepackage{mathptmx}
\usepackage{bm}
\usepackage{times}

\begin{document}


\title{Trans-Planckian enhancements of the primordial non-Gaussianities}

\author{Hael Collins}
\email{hael@nbi.dk}
\affiliation{The Niels Bohr International Academy, The Niels Bohr Institute, 2100 Copenhagen \O, Denmark}
\author{R.~Holman}
\email{rh4a@andrew.cmu.edu}
\affiliation{Department of Physics, Carnegie Mellon University, 
Pittsburgh, Pennsylvania\ \ 15213}

\date{\today}

\begin{abstract}
This article examines how breaking a Lorentz-invariant description of nature at tiny space-time intervals would affect the non-Gaussian character of the pattern of primordial perturbations left by inflation.  We specifically study a set of irrelevant operators that preserve the spatial symmetries of the usual inflationary background.  The non-Gaussian component in the primordial fluctuations can be much larger than the usual, small, inflationary prediction and can thus lead to much stronger constraints on the role of ``trans-Planckian'' physics in inflation than those from the measurements of the primordial power spectrum.
\end{abstract}


\maketitle

\section{Introduction}
\label{introduce}

The idea of inflation \cite{texts}---that the universe underwent a burst of extreme expansion during its infancy---leads to several clear predictions about how it should appear to us today.  One of the most appealing of these predictions is that the universe inevitably must have had some spatial variability.  While the observed relics of the earlier universe show that it was a far more uniform, homogeneous place in its youth than it is today, they also show that it does appear to have always contained at least some small spatial fluctuations from the earliest times.

In observing our universe, what we can see directly is the variation in how some particular material is distributed, such as the variations in the density or temperature of the plasma that produced the cosmic microwave background radiation \cite{cmb} or in how clusters of galaxies are distributed over vast regions of space \cite{sdss}.  From these measurements, we try to infer the pattern of the yet more primitive {\it primordial perturbations\/}, the primeval fluctuations in the space-time background itself.  Inflation gives a simple picture for generating these primordial perturbations, and the predicted pattern so far agrees with what is inferred from what we see \cite{wmap}.

In this picture, these fluctuations arise through a purely quantum process.  A new field is typically introduced that is responsible for the period of inflation.  Both this field and the space-time metric are then treated as a classical, spatially uniform part plus a small quantum mechanical piece.  By itself, introducing a quantum piece would not be enough to produce the observed pattern of fluctuations.  For example, in a universe that expands at a decelerating rate, the size of the region we can observe grows faster than the range of causal processes.  However, inflation reverses this ordering by proposing that for a time the universe expanded at an accelerating rate.  Information---in this case, the pattern of the quantum fluctuations---is stretched until it lies beyond the reach of any {\it subsequent\/} causal process for as long as the accelerating expansion continues.  By the end of inflation, the size of a causally connected region of the universe can then be much larger than the tiny part of it that can be seen by a single observer.

Despite the attractiveness of this idea, it still has many deep problems, which occur fairly independently of how it is implemented \cite{problems}.  For example, many of the parameters in any particular model of inflation must be chosen extremely carefully to produce enough inflation; and the identity of the field responsible for inflation, and its role in a fuller physical picture of nature, is unknown.  But perhaps the most intriguing question is how inflation eludes its {\it trans-Planckian problem\/}.

Inflation relies on having both quantum fluctuations of gravity and an expanding background.  At low energies, and in flat space, there is a good reason for why what occurs at tiny scales, such as the scale at which the interactions of quantum gravity should become strong (called the Planck scale), ought to have little effect on what happens at much larger distances.  In an expanding background, however, this clean separation of scales no longer holds.  If we simply wait long enough, a quantum fluctuation the size of a Planck length or smaller will be stretched to an enormous size and frozen into the background, until much later when it comes to influence some observable feature of our universe.  This question of whether and the extent to which the details of nature at such extremely small scales actually influence any observable feature of our universe corresponds to the trans-Planckian problem of inflation \cite{problems}.

This problem has so far not been fully solved\footnote{In speaking of the trans-Planckian problem, we mean the form in which it was originally described by Brandenberger \cite{problems}, and not what should be the characteristic size of the ``trans-Planckian'' (or more accurately, the ``trans-Hubble'') corrections in a particular scenario:  that is, a choice for the Largrangian {\it and\/} the quantum state for the fluctuations.}.  What is usually done is to derive the signal from the trans-Planckian features of a particular scenario \cite{brandenberger,tp0,tp1,cliff,tp2,tp3,tp4,tp5,broken,effstate}.  This approach must make an assumption about the structure of nature---most especially, the structure of the quantum state of the fluctuations down to arbitrarily small scales.  In any of these scenarios, it is then possible to describe how large, and in what form, the effects of these trans-Planckian details would be in a cosmological measurement.  But this technique does not quite answer the question of whether the universe can only be in particular states during inflation or whether some deeper principle is at work that hides any such trans-Planckian information.

The detailed pattern of the primordial perturbations could be complicated in principle; but in the standard inflationary picture, this pattern should be quite simple.  The full pattern is often described in terms of $n$-point correlation functions, which tell the extent to which the fluctuations in $n$ different places happen to be related to each other.  Since this means simultaneously comparing $n$ separate fluctuations, each of which is itself small, only the first few of these functions are ever likely to be measurable.

So far only the two-point function has been observed, but a crucial test for inflation will be the form of the three-point function.  This is because inflation predicts a highly Gaussian pattern.  In a perfectly Gaussian pattern, all even-point functions can be written in terms of the two-point function and all odd-point functions vanish.  Even though the three-point function is not expected exactly to vanish in a realistic inflationary model, it is still suppressed, even beyond what would have been expected from the inherent smallness of the fluctuations \cite{nongauss,maldacena}.  Therefore, measuring the smaller three-point function is often a promising place to look for signals of something outside the standard inflationary picture since they would stand out all the more prominently \cite{3point}.

This article examines the contributions from a particular set of trans-Planckian effects to the three-point function of the primordial perturbations.  The scenario that we shall be considering is one where the symmetry between the spatial and the time directions is explicitly broken \cite{broken}.  To keep the analysis otherwise as general as possible, we shall describe this breaking by using an effective theory language, rather than by implementing it through a specific model.  This approach can thus be applied to a range of models---for example, one with an excited field that interacts with the quantum fluctuations of the background, or a model that breaks the usual flat-space symmetries through quantum gravitational effects, among other possibilities.  The approach also reproduces some of the detailed signals that occur when we modify the quantum state of the fluctuations---rather than how they evolve over time---but in a more familiar language and setting \cite{effstate}.

We shall find in the course of this work that the observational constraints on the three-point function for the primordial perturbations provide much better limits on the possibilities for certain trans-Planckian effects---at least for those breaking some of the space-time symmetries as discussed here.  Before reaching this point, we first shall review how inflation produces the primordial fluctuations and also define exactly what we mean by the trans-Planckian problem.  Once this review is done, we shall introduce an effective description for the symmetry-breaking and evaluate the three-point function for this theory.

\section{The trans-Planckian problem reviewed}
\label{plank}

The trans-Planckian problem is more or less inherent in any inflationary universe that has just a bit more than the minimal amount of inflation necessary to address the horizon problem.\footnote{the number of extra ``{\it e\/}-folds'' of inflation needed being the natural logarithm of the ratio between the Planck mass and the inflationary expansion rate}  We shall illustrate it here for the simplest inflationary model, where there is only one scalar field $\phi$ called the {\it inflaton\/} directly participating in the inflation.  If this field has a potential $V(\phi)$, then the action determining the inflationary dynamics is 
\begin{equation}
S = \int d^4x\, \sqrt{-g} \Bigl\{ 
{\textstyle{1\over 2}} M_{\rm pl}^2\, R 
+ {\textstyle{1\over 2}} g^{\mu\nu} \partial_\mu\phi \partial_\nu\phi - V(\phi) \Bigr\} ,
\label{action}
\end{equation}
where the (reduced) Planck mass $M_{\rm pl}$ is defined in terms of Newton's constant $G$ through
\begin{equation}
M_{\rm pl} = {1\over\sqrt{8\pi G}} . 
\label{Mpldef}
\end{equation}

The background that is typically assumed for inflation, based upon the observed properties of the early universe, is one with almost no spatial variation at any particular time.  Neglecting this variation, a suitable definition of the time coordinate, $\eta$, allows this background to be described by the metric 
\begin{equation}
ds^2 = a^2(\eta)\, \eta_{\mu\nu}\, dx^\mu dx^\nu 
= a^2(\eta)\, \bigl[ d\eta^2 - d\vec x\cdot d\vec x \bigl] . 
\label{metric}
\end{equation}
Of course, this background by itself cannot completely describe the universe, even at early times, since the universe is never perfectly featureless.  In the inflationary picture, spatial variations are hypothesized to have arisen from random quantum fluctuations of both the inflaton $\phi$ and the metric $g_{\mu\nu}$ about spatially invariant values,
\begin{eqnarray}
\phi(\eta,\vec x) 
&\!\!\!=\!\!\!& \phi_0(\eta) + \delta\phi(\eta,\vec x) 
\nonumber \\
g_{\mu\nu}(\eta,\vec x) 
&\!\!\!=\!\!\!& a^2(\eta)\eta_{\mu\nu} + \delta g_{\mu\nu}(\eta,\vec x) . 
\label{perts}
\end{eqnarray}
Here, $\phi_0(\eta)$ and $a^2(\eta)\eta_{\mu\nu}$ are essentially classical, while both $\delta\phi$ and $\delta g_{\mu\nu}$ are the quantum fluctuations.  Notice already that since inflation depends on the quantum fluctuations of the metric---essentially quantum gravity---it remains predictive\footnote{since otherwise we ought to have included higher order curvature terms in the action.  At Planckian scales, all such terms can become equally important---at least in principle.} only so long as the characteristic size of a fluctuation is large compared to the Planck length, $1/M_{\rm pl}$.

The metric fluctuation, $\delta g_{\mu\nu}$, would appear to contain ten independent functions.  Since the background, as well as the operators that we shall consider later, preserve the spatial symmetries, we can characterize these ten by how they transform with respect to this unbroken spatial symmetry.  Four of them transform as scalar fields, four as the components of three-vectors, and the last two as the components of a two-index tensor, or gravity waves.  Only the scalar fluctuations have been observed thus far, so we shall concentrate solely on them here.  Of the four scalar fields, a few are redundant, being associated with a particular choice of the coordinates, so they carry no independent physical information.  As a result, we can write a general set of scalar fluctuations of the metric in the following form, 
\begin{eqnarray}
&&\!\!\!\!\!\!\!\!\!\!\!\!\!\!\!\!\!\!\!
\delta g_{\mu\nu}(\eta,\vec x)\, dx^\mu dx^\nu 
\nonumber \\
&\!\!\!=\!\!\!& 2a^2(\eta) \Bigl\{ 
\bigl( \Phi - a^{-1} [a(B-E')]'\bigr)\, d\eta^2 - \partial_i B\, d\eta dx^i 
\nonumber \\
&&\qquad
+\ \bigl( (\Psi +aH(B-E'))\delta_{ij}-\partial_i\partial_j E \bigr)\, dx^idx^j 
\Bigr\} . \quad
\label{scalargmunu}
\end{eqnarray}
In this equation, a prime represents a derivative with respect to the conformal time $\eta$, and $H$ is the natural dynamical scale given by the expansion rate of the universe, 
\begin{equation}
H = {a'\over a^2} . 
\label{Hdef}
\end{equation}
The fields $E(\eta,\vec x)$ and $B(\eta,\vec x)$ have no physical effect, at least when the action is expanded to second order in the fluctuations, and so we are left with just $\Psi(\eta,\vec x)$ and $\Phi(\eta,\vec x)$.  These fields were originally introduced by Bardeen \cite{bardeen}.  The reason for the multiple appearances of $B-E'$ in $\delta g_{\mu\nu}$ is that in this form the fields $\Psi$ and $\Phi$ remains unaltered even if we redefine our coordinates by a small amount.  Similarly, we can define the quantum part of the inflaton, 
\begin{equation}
\delta\phi = \delta\varphi - \phi'_0(B-E') , 
\label{dvarphidef}
\end{equation}
so that $\delta\varphi(\eta,\vec x)$ is also unchanged for a tiny coordinate transformation.

We shall describe the evolution of this quantum field theory in the interaction picture.  There, the evolution of operators, such as a product of these fields, is given by the free, or quadratic, part of the Lagrangian, while the evolution of the states in which they are evaluated is determined by the interacting part.  Obtaining even just the free part of the action in terms of $\delta\varphi$, $\Psi$, $\Phi$, $E$, and $B$ requires a rather long calculation \cite{perts}, so we only mention the final result here.

In the course of this calculation, we encounter one constraint among the physical fields, 
\begin{equation}
aH\Phi + \Psi' = 4\pi G \phi'_0 \delta\varphi ,
\label{constraint}
\end{equation}
which, when imposed, allows us to write the {\it dynamical\/} part of the action in terms of but a single linear combination of the fields,
\begin{equation}
\varphi = \delta\varphi + {\phi'_0\over aH} \Psi
\label{varphidef}
\end{equation}
as 
\begin{equation}
S_0[\varphi] = \int d^4x\, a^2 \Bigl\{ 
{\textstyle{1\over 2}} \eta^{\mu\nu} \partial_\nu\varphi\partial_\nu\varphi 
- {\textstyle{1\over 2}} a^2 m^2 \varphi^2 
\Bigr\} . 
\label{freeaction}
\end{equation}
There are also many further terms for the other fields, but they are all total derivatives.  The effective mass $m$ of the field $\varphi$ is not constant.  It is partially determined by the second derivative of the potential, as might be expected, but it also receives contributions from the background as well,
\begin{equation}
m^2 = {\delta^2 V\over \delta\phi_0^2} + {4\over a} H' 
+ {2\over a^2} {H^{\prime\prime}\over H} 
- {2\over a^2} {H^{\prime 2}\over H^2} . 
\label{littlemdef}
\end{equation}

We can now use this free action to solve for how the field evolves.  We first expand the field in terms of its individual eigenmodes, $\varphi_k(\eta)$, 
\begin{equation}
\varphi(\eta,\vec x) = \int {d^3\vec k\over (2\pi)^3}\, 
\bigl\{ \varphi_k(\eta) e^{i\vec k\cdot\vec x} a_{\vec k}
+ \varphi_k^*(\eta) e^{-i\vec k\cdot\vec x} a_{\vec k}^\dagger \bigr\} . 
\label{phiops}
\end{equation}
From the quadratic action, these modes satisfy the second-order equation
\begin{equation}
\varphi_k^{\prime\prime} + 2aH\, \varphi_k^\prime 
+ \bigl( k^2 + a^2m^2 \bigr) \varphi_k = 0 .
\label{KGeqn}
\end{equation}

To go any further, at least analytically, we need to make some additional assumptions about how the universe evolves.  The classical part of the inflaton, $\phi_0(\eta)$, is meant to produce the stage of accelerated expansion.  For a scalar field, this occurs when the kinetic energy is smaller than the potential energy; moreover, this phase must last long enough to produce a sufficient amount of expansion to circumvent the horizon problem.  If we introduce some dimensionless parameters, 
\begin{equation}
\epsilon \equiv - {H'\over aH^2}, \qquad 
\delta + 1 \equiv {1\over aH} {\phi_0^{\prime\prime}\over\phi_0^\prime}, \qquad 
\xi \equiv {\epsilon'+\delta'\over aH} ,
\label{slowly}
\end{equation}
then these requirements on how the classical parts of the field and the background evolve are
\begin{equation}
\epsilon \ll 1, \qquad 
\delta \ll 1,  
\label{slowlyrolly}
\end{equation}
and 
\begin{equation}
\xi,\epsilon',\delta' 
\sim {\cal O}\bigl(\epsilon^2,\delta^2,\epsilon\delta \bigr) . 
\label{noxi}
\end{equation}

Without any approximations, the effective mass of the field in terms of these parameters is 
\begin{equation}
m^2 = - H^2 \bigl[ (3+\delta)(\epsilon+\delta) + \xi \bigr] ; 
\label{mexact}
\end{equation}
but in the limit where we neglect higher order corrections in these parameters, we can more simply use 
\begin{equation}
m^2 \approx - 3H^2(\epsilon+\delta) , 
\qquad
aH \approx - {1+\epsilon\over\eta} , 
\label{mapprox}
\end{equation}
treating $\epsilon$ and $\delta$ as more or less constant.  The differential equation for the modes then becomes a Bessel equation, 
\begin{equation}
\varphi_k^{\prime\prime} - {2(1+\epsilon)\over\eta} \varphi_k^\prime 
+ \biggl[ k^2 - {3(\epsilon+\delta)\over\eta^2} \biggr] \varphi_k \approx 0 .
\label{Besseleqn}
\end{equation}
Being a second-order equation, its solution has two free constants of integration.  One of these, the normalization of the field, is fixed by the equal-time commutation relation between the field and its momentum, 
\begin{equation}
\bigl[ \varphi(\eta,\vec x), \pi(\eta,\vec y) \bigr] 
= i\delta^3(\vec x-\vec y),
\qquad\quad \pi = a^2\varphi' ,
\label{cancom}
\end{equation}
which becomes 
\begin{equation}
a^2 \bigl[ \varphi_k \varphi_k^{\prime *} - \varphi_k^*\varphi'_k \bigr] = i .
\label{Wronk}
\end{equation}

So far, the trans-Planckian problem has not appeared at all---at rather, it has not appeared explicitly.  To determine the time-evolution of the field fully requires one further condition.  What is usually done is to choose the state which matches with the Minkowski vacuum at short distances \cite{bunch}; this condition, plus the normalization constraint, determines the complete form of the modes,
\begin{equation}
\varphi_k(\eta) 
= - {i\sqrt{\pi}\over 2} {H(-\eta)^{3/2}\over 1+\epsilon} H_\nu^{(1)}(-k\eta) , 
\label{modescomplete}
\end{equation}
and consequently the time-evolution of the field.  Here $H_\nu^{(1)}(-k\eta)$ is a Hankel function whose index, to leading order in the slowly rolling parameters, is 
\begin{equation}
\nu = {3\over 2} + 2\epsilon + \delta . 
\label{nuapprox}
\end{equation}

The only trouble in this reasoning, the last condition especially, is that the dramatic expansion that occurs during inflation can easily stretch fluctuations whose characteristic scale is the size of a Planck length---or even smaller---to a size that has an observable effect on the universe at large scales.  Our whole derivation has been based on a simple form for the gravitational action, $\int d^4x\, \sqrt{-g} R$, where higher order terms---powers of the Riemann, Ricci or scalar curvature tensors---are suppressed by an appropriate number of powers of the Planck length, $1/M_{\rm pl}$.  However, when the physical wave number of the mode, $k/a(\eta)$, is itself of the order of the Planck scale, $k/a\sim M_{\rm pl}$, then it is no longer consistent to neglect such terms and we can no longer trust the starting point that is used to derive any inflationary prediction.  At such scales, it is not even immediately obvious that space-time can be treated as a classical, flat geometry.

We can state this ``trans-Planckian'' problem a little more precisely by determining which wave numbers $k$ are both safe and relevant.  By {\it relevant\/}, we mean a mode that was inside the horizon at some point during inflation---otherwise inflation would not allow a causal explanation for these modes.  Therefore, let us define a minimally early $\eta_0$ to be the time when a mode $k_{\rm min}$, just reentering the horizon today,\footnote{Or, if the universe is indeed accelerating again, it would be the last mode to have entered the horizon before the recent acceleration began.} was just leaving the horizon during inflation,
\begin{equation}
{k_{\rm min}\over a(\eta_0)} = H(\eta_0) . 
\label{relevantk}
\end{equation}
In a slowly rolling universe, $aH \approx -(1+\epsilon)/\eta$, so this condition gives
\begin{equation}
-k_{\rm min}\eta_0 = 1 + \epsilon \sim 1 . 
\label{relevantketa}
\end{equation}
Conversely, a {\it safe\/} mode is one whose variation is larger than the Planck length---otherwise our perturbative description of quantum gravity in not applicable.  We can similarly define a maximal wave number $k_\star$ to have a physical wavelength equal to a Planck length at $\eta_0$, 
\begin{equation}
{k_\star\over a(\eta_0)} = M_{\rm pl} . 
\label{safek}
\end{equation}
Replacing $\eta_0$ with $k_{\rm min}$ we find that 
\begin{equation}
k_\star \approx {M_{\rm pl}\over H_0} k_{\rm min} , 
\label{kmaxkmin}
\end{equation}
where we have written $H_0\equiv H(\eta_0)$ and have neglected corrections suppressed by $\epsilon$.  Thus, without an assumption about the theory in the ``trans-Planckian'' regime, we can only safely apply inflationary predictions for wave numbers between the range 
\begin{equation}
k_{\rm min} < k < {M_{\rm pl}\over H_0} k_{\rm min} , 
\label{krange}
\end{equation}
that is, $\log(M_{\rm pl}/H_0)$ orders of magnitude.  Note that in this context, we are calling even choosing the state that matches the flat-space vacuum at short distances an ``assumption,'' thought it is perhaps a fairly conservative one.

This range is in fact the largest for which a perturbative description of the quantum fluctuations is {\it generally\/} applicable.  In almost any inflationary model it is bound to be narrower, or even nonexistent.  The ``initial'' time, $\eta_0$, that we introduced will not typically be the actual beginning of inflation.  Here, it was merely chosen as the latest time possible if inflation is to provide a causal explanation of the fluctuations whose effects are being observed today.

To see the problem a little more clearly, suppose that inflation began at some time $\eta_{\rm begin} < \eta_0$.  Then, since even more expansion occurred for the physically relevant modes, the trans-Planckian threshold will be lowered so that the allowed range will also be diminished, 
\begin{equation}
k_{\rm min} < k < \biggl( {M_{\rm pl}\over H_0} 
{\eta_0\over \eta_{\rm begin}} \biggr) k_{\rm min} . 
\label{smallerrange}
\end{equation}
In our coordinates, conformal times are always negative, so $\eta_0/\eta_{\rm begin} < 1$.  If we extend sufficiently far back, when $\eta_{\rm begin} \sim (M_{\rm pl}/H_0)\eta_0$, then {\it every\/} mode whose influence we are witnessing in the universe today would have been at one time smaller that the Planck length.

To avoid this problem, we shall not consider any conformal times before $\eta_0$, or equivalently, wavenumbers whose value is above $k_\star$.  Our treatment of the fluctuations will thus be only an effective one.  We shall assume that the field is in the state whose modes have the standard form
\begin{equation}
\varphi_k(\eta) = -{i\sqrt{\pi}\over 2} {H(-\eta)^{3/2}\over 1+\epsilon} 
H_\nu^{(1)}(-k\eta) ,
\label{modesagain}
\end{equation}
at least within the range $k_\star (H_0/M_{\rm pl}) < k < k_\star$, though we shall not be able to say how the universe found itself in this state in the first place.  In exchange, we can reliably treat the theory perturbatively, calculating the signals of any new effects consistently.  Here we shall examine the corrections to the three-point correlation function of the field $\varphi$ that occur when some of the space-time symmetries are explicitly broken at short distances.  Before introducing these effects, we shall first describe in a little more detail how we determine the full time-evolution of this theory.

\section{Time-evolution}
\label{evolve}

As we mentioned briefly, we shall be treating the time-evolution of the theory using the interaction picture.  In this picture, we divide the action or Lagrangian into its free and interacting parts.  The free part corresponds to the set of terms that are quadratic in the field and have a mass dimension of four or less, while the interacting part contains everything else,
\begin{equation}
S[\varphi] = S_0[\varphi] 
+ \int d\eta\, a^4(\eta) \int d^3\vec x\, {\cal L}_I[\varphi(\eta,\vec x)] . 
\label{action0I}
\end{equation}
We have just seen how the free part $S_0[\varphi]$ determines the evolution of the field.  The interacting part correspondingly determines how states evolve.  

Let us denote our choice for the initial state, implicitly defined by our choice of the modes $\varphi_k(\eta)$, by 
\begin{equation}
|0\rangle \equiv |0(\eta_0)\rangle . 
\label{vacuumdef}
\end{equation}
If we rewrite the Lagrangian as an interaction Hamiltonian, 
\begin{equation}
H_I(\eta) = - a^4(\eta) \int d^3\vec x\, {\cal L}_I[\varphi(\eta,\vec x)] , 
\label{Hint}
\end{equation}
then the time evolution of this or any other state is found by applying Dyson's equation, 
\begin{equation}
|0(\eta)\rangle = T e^{-i\int_{\eta_0}^\eta d\eta'\, H_I(\eta')} |0\rangle . 
\label{dyson}
\end{equation}

The quantum fluctuations generated by inflation are meant to provide the initial set of perturbations in the space-time for the more conventional stages of the universe, which come long after inflation has ended and the evolution is governed entirely by the matter and radiation in the universe, without any more exotic ingredients.  Since the pattern of these perturbations can be quite complex, it is usually described by saying how the perturbations in different places are correlated with each other.  Each of these {\it correlation functions\/}, or {\it n-point functions\/}, contains only a part of the total information contained in the full pattern.  So from each correlation in the density perturbations that we observe, we can correspondingly infer a correlation function of the original fluctuations produced during inflation,
\begin{equation}
\langle 0(\eta) | \varphi(\eta,\vec x_1) \cdots 
\varphi(\eta,\vec x_n) | 0(\eta)\rangle . 
\label{npoint}
\end{equation}
Since the perturbations at any one place $\varphi(\eta,\vec x)$ are small, looking at the fluctuations in several places at once, as in a higher $n$-point function, becomes progressively more difficult to measure in practice.  So far only the two-point function has been measured, but, depending upon its actual size, experimental precisions might soon be sufficiently good to see three-point function as well.

For a Gaussian pattern of random fluctuations, the entire pattern is completely determined by the two-point function:  odd-point functions vanish and even-point functions can be expressed as a series of products of two-point functions.  In the interaction picture, any non-Gaussian features in the pattern of the fluctuations must necessarily come entirely from the evolution of the state.  Thus, in the example of a three-point function, the expectation value of three fields vanishes in the initial state,
\begin{equation}
\langle 0 | \varphi(\eta, \vec x)\varphi(\eta, \vec y)\varphi(\eta, \vec z) | 0\rangle = 0 ,
\label{no3vac}
\end{equation}
so the leading terms come from extracting one factor of the interaction Hamiltonian in Dyson's equation,\footnote{In this article we shall implicitly be using the Schwinger-Keldysh \cite{sk} formalism to find the time-evolution of the matrix elements, which is also described in \cite{effstate} in the form that we shall be applying it.  However, the notation in the following equation might be more familiar \cite{wein}, and at this order there are no ambiguities \cite{badinin} in how to evolve a matrix element.} 
\begin{eqnarray}
&&\!\!\!\!\!\!\!\!\!\!\!\!\!\!\!\!\!
\langle 0(\eta)|\varphi(\eta,\vec x) \varphi(\eta,\vec y) \varphi(\eta,\vec z) 
| 0(\eta)\rangle
\nonumber \\
&=&
i \int_{\eta_0}^\eta d\eta'\, 
\langle 0| \bigl[ H_I(\eta'), 
\varphi(\eta,\vec x) \varphi(\eta,\vec y) \varphi(\eta,\vec z) \bigr] 
| 0\rangle 
+ \cdots .
\nonumber \\
&& 
\label{time3point}
\end{eqnarray}
The largest effect is accordingly produced by the cubic terms in the field $\varphi$.  

Cubic terms are already present in the standard inflationary picture, and they can be found by expanding the original action for gravity and the inflaton to third order in the perturbations $\delta g_{\mu\nu}(\eta,\vec x)$ and $\delta\varphi(\eta,\vec x)$.  These terms and their contribution to the three-point function have been calculated elsewhere \cite{nongauss,maldacena}, so here we only mention the natural size of their role in three-point function in the slowly rolling limit of inflation so that we can compare whether the symmetry-breaking effects that we shall introduce in the next section are larger or smaller than those already present in the usual description of inflation.

Before doing so, we should first convert from the field $\varphi$, which was defined to have the canonical normalization for its kinetic term---appropriate for treating the fluctuations as a quantum field theory---back to a normalization appropriate for cosmological applications.  Since the basic purpose of inflation is to generate a set of random fluctuations in the curvature of the background, we shall instead use a field $\zeta$ defined by rescaling $\varphi$ as follows,
\begin{equation}
\zeta = {aH\over\phi'_0} \varphi = \Psi + {aH\over\phi'_0} \delta\varphi . 
\label{zetadef}
\end{equation}
The fact that $\zeta=\Psi+\cdots$ means that this field behaves like the three-dimensional curvature along a constant-time, spatial surface, especially once a particular mode has been stretched well outside the horizon, $k/a \ll H$.  The rescaling factor $aH/\phi'_0$ can be expressed in a slightly more useful form by applying the following relation derived from the $\eta\eta$ component of the background Einstein equation and rewritten in terms of the slow-roll parameter $\epsilon$,
\begin{equation}
\biggl( {\phi'_0\over aH} \biggr)^2 = {\epsilon\over 4\pi G} 
= {1\over 2} M_{\rm pl}^2\epsilon .
\label{redefeps}
\end{equation}
Substituting in the expression for $\zeta$ into the quadratic action for $\varphi$ yields, 
\begin{equation}
S_0[\zeta] = \int d^4x\, a^2 
\bigl( {\textstyle{1\over 2}} M_{\rm pl}^2\epsilon \bigr)
\bigl\{ {\textstyle{1\over 2}} \eta^{\mu\nu} 
\partial_\mu\zeta \partial_\nu\zeta \bigr\} , 
\label{zetaaction}
\end{equation}
which shows even more clearly that the field $\zeta$ only remains dynamical as long as the de Sitter symmetry is broken, $\epsilon\not= 0$. 

We can make a standard, though extremely crude, estimate for the size of the three-point function.  Our purpose with this estimate is not to evaluate the three-point function accurately, but rather is to have a rough idea of the size of the standard non-Gaussianities to compare with those that we shall investigate in the next section.  To do so, we shall treat $\zeta$ as though it consisted of an entirely Gaussian part $\zeta^{(g)}$ plus a small non-Gaussian piece scaling as $f_{\rm nl}\zeta^{(g)\, 2}$, 
\begin{equation}
\zeta \approx \zeta^{(g)} + f_{\rm nl} \bigl( \zeta^{(g)} \bigr)^2 . 
\label{zetagdef}
\end{equation}
The theory is scale invariant in the de Sitter limit, so we shall assume that $f_{\rm nl}$ is of the same order as the slow-roll parameters, $f_{\rm nl}\sim\epsilon,\delta$, since $f_{\rm nl}$ is associated with the breaking of the scale-invariance.\footnote{A more careful treatment shows that this is the case \cite{maldacena}.}  The modes $\zeta_k^{(g)}(\eta)$ for this field can be obtained by taking the de Sitter limit ($\epsilon,\delta\to 0$) of our expression $\varphi_k(\eta)$ at the end of the last section, and then rescaling in the way just described, 
\begin{equation}
\zeta_k^{(g)}(\eta) = {i\over\sqrt{\epsilon}} {H\over M_{\rm pl}} 
{i-k\eta\over k^{3/2}} e^{-ik\eta} . 
\label{zetagmodes}
\end{equation}

The small quadratic piece in the field means that a tree-level evaluation---one ignoring the time-evolution of the state, $|0(\eta)\rangle$---now has a nonvanishing value,
\begin{eqnarray}
&&\!\!\!\!\!\!\!\!\!\!\!\!\!\!
\langle 0 | \zeta(\eta,\vec x)\zeta(\eta,\vec y)\zeta(\eta,\vec y) |0\rangle 
\nonumber \\
&\!\!\!\approx\!\!\!& 
f_{\rm nl}\, \langle 0 | \zeta^{(g)}(\eta,\vec x) \zeta_g^{(g)}(\eta,\vec x) 
\zeta_g^{(g)}(\eta,\vec y) \zeta_g^{(g)}(\eta,\vec z) |0\rangle 
\nonumber \\
&& 
+ f_{\rm nl}\, \langle 0 | \zeta^{(g)}(\eta,\vec x) \zeta_g^{(g)}(\eta,\vec y) 
\zeta_g^{(g)}(\eta,\vec y) \zeta_g^{(g)}(\eta,\vec z) |0\rangle 
\nonumber \\
&& 
+ f_{\rm nl}\, \langle 0 | \zeta^{(g)}(\eta,\vec x) \zeta_g^{(g)}(\eta,\vec y) 
\zeta_g^{(g)}(\eta,\vec z) \zeta_g^{(g)}(\eta,\vec z) |0\rangle 
\nonumber \\
&&
+ \cdots . 
\label{NG3point}
\end{eqnarray}
Evaluating this expression for the modes that have been stretched well outside the horizon ($\eta\to 0$ compared with any of the relevant wavenumbers), yields
\begin{eqnarray}
&&\!\!\!\!\!\!\!\!\!\!\!\!\!\!\!\!\!\!\!\!\!\!
\langle 0 | \zeta(\eta,\vec x)\zeta(\eta,\vec y)\zeta(\eta,\vec y) |0\rangle 
\nonumber \\
&\!\!\!\approx\!\!\!& 
{2f\over\epsilon^2} {H^4\over M_{\rm pl}^4} 
\int {d^3\vec k_1\over (2\pi)^3} {d^3\vec k_2\over (2\pi)^3} 
{d^3\vec k_3\over (2\pi)^3}\,  
e^{i\vec k_1\cdot\vec x} e^{i\vec k_2\cdot\vec y} e^{i\vec k_3\cdot\vec z}
\nonumber \\
&&\qquad\qquad
(2\pi)^3 \delta^3(\vec k_1+\vec k_2+\vec k_3) 
{k_1^3+k_2^3+k_3^3\over k_1^3 k_2^3 k_3^3}
\nonumber \\
&&
+ {f\over\epsilon^2} {H^4\over M_{\rm pl}^4} 
\int {d^3\vec k_1\over (2\pi)^3} {d^3\vec k_2\over (2\pi)^3}\,  
{1\over k_1^3 k_2^3}
\nonumber \\
&&\qquad\qquad
\bigl\{ 
e^{i\vec k_1\cdot(\vec x-\vec y)} + e^{i\vec k_1\cdot(\vec y-\vec z)} 
+ e^{i\vec k_1\cdot(\vec z-\vec x)} \bigr\}
\nonumber \\
&& 
+ \cdots 
\label{NG3pointasym}
\end{eqnarray}
The detailed dependence of the integrands on the wavenumbers is certainly not to be trusted; however, the dependence on the fixed parameters of the model---$H$, $M_{\rm pl}$, $\epsilon$---should be more or less correct.  What we have learned then is that the typical size for the three-point function is 
\begin{equation}
\langle 0(\eta) | \zeta(\eta,\vec x)\zeta(\eta,\vec y)\zeta(\eta,\vec y) |0(\eta)\rangle 
\sim {1\over\epsilon} {H^4\over M_{\rm pl}^4} , 
\label{3pointscaling}
\end{equation}
where for simplicity we have approximated $f\sim\epsilon$.  It is important later to remember this scaling; any new effects smaller than this size would be unobservable, compared those that are already expected to be present in most inflationary models.

\section{The three-point function}
\label{threepoint}

We shall now consider how a particular set of new effects, generated at very short distances, would appear in the three-point function of the primordial perturbations.  To have a fairly general picture and yet one that is at the same time still compatible with the usual background assumed for inflation, we shall examine operators that break the symmetry between the time and the space dimensions while still keeping the invariance under spatial rotations and translations.  We shall assume that such symmetry-breaking operators are generated by some new physical property of nature with an associated energy scale $M$, which is larger than the inflationary scale $H$ but not necessarily equal to the Planck scale.

The leading irrelevant operators that directly affect the three-point function are the cubic ones in the field $\varphi$, 
\begin{equation}
{\cal L}_I = {1\over 6} {d_1\over M}\, H^2\varphi^3 
- {1\over 2} {d_2\over a^2M}\, \varphi^2 \vec\nabla\cdot\vec\nabla\varphi 
+ \cdots . 
\label{leadirrel}
\end{equation}
Here the operators have been written in a form appropriate for the conformally flat background.  We could also have expressed them in a more general form, applicable to any background, by introducing a time-like unit vector, $n^\mu$, with $n_\lambda n^\lambda = 1$.  The surfaces orthogonal to this vector have a naturally induced three-dimensional metric given by
\begin{equation}
h_{\mu\nu} = g_{\mu\nu} - n_\mu n_\nu , 
\label{induced}
\end{equation}
and the curvature $K_{\mu\nu}$ associated with how these surfaces are embedded in the larger space,
\begin{equation}
K_{\mu\nu} = h_\mu^{\ \lambda} \nabla_\lambda n_\nu , 
\label{extrinsic}
\end{equation} 
is called the extrinsic curvature.  For our conformally flat background, we have 
\begin{eqnarray}
n_\mu\, dx^\mu &\!\!\!=\!\!\!& a\, d\eta 
\nonumber \\
h_{\mu\nu}\, dx^\mu dx^\nu &\!\!\!=\!\!\!& - a^2\, d\vec x\cdot d\vec x 
\nonumber \\
K_{\mu\nu}\, dx^\mu dx^\nu &\!\!\!=\!\!\!& - a^2H\, d\vec x\cdot d\vec x . 
\label{confcases}
\end{eqnarray}
In terms of these tensors, we can express our interaction Lagrangian in a slightly more general form,
\begin{equation}
{\cal L}_I = {1\over 54}{d_1\over M}\, K^2\varphi^3 
+ {1\over 2}{d_2\over M}\, \varphi^2 
\bigl[ h^{\mu\nu}\nabla_\mu\nabla_\nu - K n^\lambda \nabla_\lambda \bigr] \varphi 
+ \cdots , 
\label{leadirrelgen}
\end{equation}
where $K=g^{\mu\nu}K_{\mu\nu}$.

Returning to the conformally flat frame, the interaction Hamiltonian for these operators is then
\begin{equation}
H_I(\eta) = - {1\over M} \int d^3\vec x\, \Bigl\{ 
{1\over 6}\, d_1 a^4H^2 \varphi^3 
- {1\over 2}\, d_2 a^2\varphi^2 \vec\nabla\cdot\vec\nabla\varphi 
\Bigr\} . 
\label{}
\end{equation}
The leading order contribution to the three-point function of the field $\varphi(\eta,\vec x)$ comes entirely from the evolution of the state, which is 
\begin{eqnarray}
&&\!\!\!\!\!\!\!\!\!\!\!
\langle 0(\eta)|\varphi(\eta,\vec x) \varphi(\eta,\vec y) \varphi(\eta,\vec z) 
| 0(\eta)\rangle
\nonumber \\
&\!\!\!=\!\!\!& 
\int {d^3\vec k_1\over(2\pi)^3} {d^3\vec k_2\over(2\pi)^3} 
{d^3\vec k_3\over(2\pi)^3}\, 
e^{i\vec k_1\cdot\vec x} e^{i\vec k_2\cdot\vec y} e^{i\vec k_3\cdot\vec z} 
(2\pi)^3 \delta^3(\vec k_1 + \vec k_2 + \vec k_3) 
\nonumber \\
&&\quad
{1\over MH^2} \int_{\eta_0}^\eta d\eta'\, 
\biggl\{ {d_1\over\eta^{\prime 4}} 
+ {d_2\over\eta^{\prime 2}} (k_1^2+k_2^2+k_3^2) \biggr\} 
\nonumber \\
&&\qquad\qquad
i \bigl\{ 
\varphi_{k_1}(\eta) \varphi_{k_2}(\eta) \varphi_{k_3}(\eta)
\varphi_{k_1}^*(\eta') \varphi_{k_2}^*(\eta') \varphi_{k_3}^*(\eta')
\nonumber \\
&&\qquad\qquad
- 
\varphi_{k_1}^*(\eta) \varphi_{k_2}^*(\eta) \varphi_{k_3}^*(\eta)
\varphi_{k_1}(\eta') \varphi_{k_2}(\eta') \varphi_{k_3}(\eta')
\bigr\}
\nonumber \\
&&
+ \cdots . 
\label{leading3point}
\end{eqnarray}
Since these effects only have a significance when they are at least as large as those already present in the usual inflationary picture, it is only necessary to keep the leading terms in the slowly rolling limit.  Thus we can set $\epsilon = \delta = 0$ in the general expression for the mode functions $\varphi_k(\eta)$; the index of the Hankel functions then becomes $\nu={3\over 2}$.  For that value, the expression for the mode functions simplifies considerably,
\begin{equation}
\varphi_k(\eta) \approx 
{H\over\sqrt{2k^3}} (i-k\eta) e^{-ik\eta} 
+ {\cal O}(\epsilon,\delta) . 
\label{modes3halves}
\end{equation}
This form is the familiar one for a de Sitter background.

Even with this simplification, the integrals associated with the time-evolution from the ``initial'' time $\eta_0$ to $\eta$ will be quite lengthy.  Fortunately, the physical setting that we are considering has several simplifying limits inherent to it.  The modes that are relevant for cosmological observations will always have been stretched well outside the horizon by the end of inflation.  In terms of the wavenumbers, this limit corresponds to sending 
\begin{equation}
-k_i\eta \to 0 \qquad i=1,2,3.
\label{latelimit}
\end{equation}
At the opposite extreme, the relevant wavenumbers are those lying within the interval, 
\begin{equation}
1 \le -k_i\eta_0 \le {M\over H_0} \qquad i=1,2,3,
\label{earlylimit}
\end{equation}
except that here we have replaced the trans-Planckian threshold that appeared in our first encounter with this expression with the scale $M$ for whatever phenomena breaks the symmetries; most conservatively, $M=M_{\rm pl}$, but it could be smaller.  

So in general, $-k_i\eta_0$ can be large and $-k_i\eta$ will always be small.  Applying both of these limits to the leading-order expression for the three-point function above, and using the de Sitter form for the modes, $\varphi_k$, we have 
\begin{widetext}
\begin{eqnarray}
\langle 0(\eta)|\varphi(\eta,\vec x) \varphi(\eta,\vec y) \varphi(\eta,\vec z) 
| 0(\eta)\rangle
&\!\!\!=\!\!\!& 
\int {d^3\vec k_1\over(2\pi)^3} {d^3\vec k_2\over(2\pi)^3} 
{d^3\vec k_3\over(2\pi)^3}\, 
e^{i\vec k_1\cdot\vec x} e^{i\vec k_2\cdot\vec y} e^{i\vec k_3\cdot\vec z} 
(2\pi)^3 \delta^3(\vec k_1 + \vec k_2 + \vec k_3) 
{k_1^3+k_2^3+k_3^3\over k_1^3k_2^3k_3^3} 
\nonumber \\
&&
{H^4\over M} \biggl\{ 
{d_1\over 12} \Bigl\{ 
{(k_1+k_2+k_3) (k_1^2+k_2^2+k_3^2) - k_1k_2k_3\over k_1^3+k_2^3+k_3^3}
- \ln\bigl|(k_1+k_2+k_3)\eta\bigr| - \gamma 
\Bigr\}
\nonumber \\
&&\qquad
+ {d_2\over 4} 
{k_1k_2k_3(k_1^2+k_2^2+k_3^2)\over (k_1+k_2+k_3)(k_1^3+k_2^3+k_3^3)} 
\eta_0 \sin\bigl[ (k_1+k_2+k_3)\eta_0 \bigr]
\nonumber \\
&&\qquad
+ {\cal O}\bigl( (k_i\eta_0)^{-1} \bigr) 
+ {\cal O}\bigl( (k_i\eta)^2 \bigr) 
\biggr\} . 
\label{corrections}
\end{eqnarray}
\end{widetext}
Notice that the second contribution in the three-point function depends on $\eta_0$, even in the limit where $k_i\eta_0\to -\infty$.  This dependence on the initial time can be less intuitive, so let us rewrite it instead as a maximal physical wavenumber as in Eq.~(\ref{safek}), except that we should now use the scale $M$ rather than the Planck mass as the cut-off for our effective treatment,
\begin{equation}
{k_\star\over a(\eta_0)} \equiv M 
\quad\Rightarrow\quad 
\eta_0 \approx - {1\over k_\star} {M\over H} . 
\label{kstarredef}
\end{equation}
The fact that this correction diverges as $k_i/k_\star\to\infty$ does not necessarily imply that the physical three-point function also diverges.  Rather, what is happening is that we are leaving the range of momenta for which our effective theory is applicable.  It is not consistent in this case to keep only the leading irrelevant operators, such as $\varphi^2\vec\nabla\cdot\vec\nabla\varphi$, since at these scales, anything of the form $\varphi^2 (\vec\nabla\cdot\vec\nabla)^d \varphi$ could give a comparable correction.

We can now transform this result into a statement about the size and shape---that is, its dependence on the wavenumbers $k_i$---of the three-point functions for the cosmologically relevant field $\zeta$.  We do so by rescaling the field $\varphi$ to obtain $\zeta$, 
\begin{equation}
\varphi(\eta,\vec x)=M_{\rm pl} \sqrt{{\epsilon\over 2}}\, \zeta(\eta,\vec x) .
\label{varphitozeta}
\end{equation}
To isolate how the size and the momentum dependence of these corrections behave, we define two dimensionless ``amplitudes'' ${\cal A}_i(k_1,k_2,k_3)$ for each of the operators by 
\begin{eqnarray}
&&\!\!\!\!\!\!\!\!\!\!\!\!\!\!\!\!\!\!\!
\langle 0(\eta)|\zeta(\eta,\vec x) \zeta(\eta,\vec y) \zeta(\eta,\vec z) 
| 0(\eta)\rangle
\nonumber \\
&\!\!\!=\!\!\!& 
\int {d^3\vec k_1\over(2\pi)^3} {d^3\vec k_2\over(2\pi)^3} 
{d^3\vec k_3\over(2\pi)^3}\, 
e^{i\vec k_1\cdot\vec x} e^{i\vec k_2\cdot\vec y} e^{i\vec k_3\cdot\vec z} 
\nonumber \\
&&\quad
(2\pi)^3 \delta^3(\vec k_1 + \vec k_2 + \vec k_3) 
{k_1^3+k_2^3+k_3^3\over k_1^3k_2^3k_3^3} {\cal A}_i(k_1,k_2,k_3)
\nonumber \\
&&
+ \cdots . 
\label{calAdef} 
\end{eqnarray}
Here ${\cal A}_1$ is associated with the operator $H^2\varphi^3$ and ${\cal A}_2$ is associated with the operator $\varphi^2 \vec\nabla\cdot\vec\nabla \varphi$.  Notice that we have extracted the same factors of the momenta as appeared in our rough estimate described in the last section.

Ultimately these primordial non-Gaussianities come to influence the later, observable properties of our universe.  Nonvanishing ${\cal A}_i$'s, for example, translate into nonvanishing values of the three-point function of temperature fluctuations in the microwave background when they are decomposed into spherical harmonic eigenmodes \cite{3ptstd}.

To leading order in the slowly rolling, $-k_i\eta_0\gg 1$ and $-k_i\eta\ll 1$ limits, we obtain the following two amplitudes for our symmetry-breaking operators,
\begin{eqnarray}
&&\!\!\!\!\!\!\!\!\!\!\!\!
{\cal A}_1(k_1,k_2,k_3) 
\label{calA1} \\
&\!\!\!=\!\!\!& 
{d_1\over 3\sqrt{2}} {1\over\epsilon^{3/2}} {H\over M} {H^3\over M_{\rm pl}^3}
\biggl\{ 
{(k_1+k_2+k_3)(k_1^2+k_2^2+k_3^2) - k_1k_2k_3\over k_1^3+k_2^3+k_3^3}
\nonumber \\
&&\qquad\qquad\qquad\qquad
- \ln\bigl|(k_1+k_2+k_3)\eta\bigr| - \gamma 
\biggr\}
\nonumber 
\end{eqnarray}
and
\begin{eqnarray}
{\cal A}_2(k_1,k_2,k_3) 
&\!\!\!=\!\!\!& 
{d_2\over\sqrt{2}} {1\over\epsilon^{3/2}} {H^3\over M_{\rm pl}^3} 
{1\over k_\star} 
{k_1k_2k_3 (k_1^2+k_2^2+k_3^2)\over (k_1+k_2+k_3)(k_1^3 + k_2^3 + k_3^3)} 
\nonumber \\
&&\times 
\sin\biggl[ {M\over H} {k_1+k_2+k_3\over k_\star} \biggr] . 
\label{calA2} 
\end{eqnarray}
The three momenta are not entirely arbitrary, but are of course conserved,
\begin{equation}
k_3 = \bigl| \vec k_1 + \vec k_2 \bigr|^{1/2} 
= \bigl( k_1^2 + k_2^2 + 2 \vec k_1\cdot\vec k_2 \bigr)^{1/2} .
\label{momcons}
\end{equation}

\begin{figure}[!tbp]
\begin{center}
\includegraphics{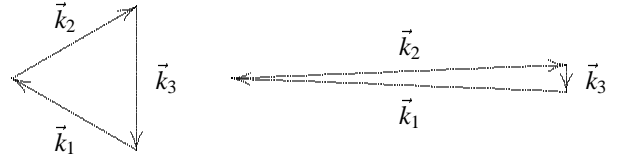}
\caption{Two representative cases for the momenta appearing in the transform of the three-point function.  In the figure on the left, all of the momenta have more or less the same magnitude, $k_1\approx k_2\approx k_3$, while the figure on the right shows the case where one of the momenta is much smaller than the other two, $k_1\approx k_2 \gg k_3$, for example.
\label{triangles}}
\end{center}
\end{figure}

To understand how these amplitudes scale in various corners of momentum space, we examine them for two representative cases.  The first is when the momenta all have similar lengths and form the sides of an equilateral triangle.  The second case corresponds to an isosceles triangle with two long sides, $\vec k_1 \approx -\vec k_2$ and one very short one, $k_3 = |\vec k_1+\vec k_2| \ll k_1,k_2$.  These cases are sketched in Fig.~\ref{triangles}.

\subsection{Case I:  The operator $H^2\varphi^3$}

The first observation is that the amplitude associated with the operator $H^2\varphi^3$ does not change very much as it ranges over the possible values of the momenta, $\{ \vec k_1, \vec k_2, \vec k_3 \}$, 
\begin{eqnarray}
{\cal A}_1^{\rm equil}(k) 
&\!\!\!\approx\!\!\!& 
{d_1\over 3\sqrt{2}} {1\over\epsilon^{3/2}} {H\over M} {H^3\over M_{\rm pl}^3} 
\biggl\{ {8\over 3} - \ln|3k\eta| - \gamma \biggr\} 
\nonumber \\ 
{\cal A}_1^{\rm iso}(k) 
&\!\!\!\approx\!\!\!& 
{d_1\over 3\sqrt{2}} {1\over\epsilon^{3/2}} {H\over M} {H^3\over M_{\rm pl}^3} 
\bigl\{ 2 - \ln|2k\eta| - \gamma \bigr\} . 
\label{A1triangle}
\end{eqnarray}
The reason for this relative insensitivity is that the amplitude ${\cal A}_1$ only depends on the momenta through sums of powers of the individual momenta, $k_1^n+k_2^n+k_3^n$, aside from one term that scales as a positive power of $k_1k_2k_3$ and which is small in the isosceles case.

To learn whether this correction is appreciable, compared with those already present in a general slowly rolling model, we divide the overall size of ${\cal A}_1$---neglecting model-dependent ($d_1$) and momentum-dependent factors---by the general estimate that we made at the end of Sec.~\ref{evolve},
\begin{equation}
\biggl( {1\over\epsilon^{3/2}} {H\over M} {H^3\over M_{\rm pl}^3} \biggr) 
\bigg/ \biggl( {1\over\epsilon} {H^4\over M_{\rm pl}^4} \biggr) 
= {1\over\sqrt{\epsilon}} {M_{\rm pl}\over M} . 
\label{ratioA1}
\end{equation}

If this operator is truly generated by ``trans-Planckian'' physics, $M\sim M_{\rm pl}$, its contribution is nearly the same as the ordinary one.  So to have an observable effect, the symmetry-breaking needs to occur at some intermediate scale, $H<M<M_{\rm pl}$.  In the standard picture, observations of the amplitude of the two-point function and bounds on the slow-roll parameter $\epsilon$ suggest an inflationary scale $H$ of the order of $10^{14}$ GeV or lower, so the upper and lower limits of this range are separated by at least five orders of magnitude.  $M_{\rm pl}/M$ could then be quite large even while $H/M$ remains small.  For a general set of irrelevant symmetry-breaking operators in this regime, the effect on the two-point function could be smaller than the usual inflationary prediction while at the same time their effect on the three-point function is much larger.  Such a result should not be too surprising since non-Gaussianities are highly suppressed in the standard picture, and here we have included an explicit cubic term.  Even though this term is nominally suppressed by $H/M$, the suppression associated with the cubic operators in the usual inflationary picture is essentially $H/M_{\rm pl}$, so that this new effect seems, comparatively speaking, quite large.

Since the value of ${\cal A}_1$ does not change much over the range of allowed momenta, its effect can be crudely described through a momentum-independent parameter such as $f_{\rm nl}$.  Stating the contribution from this operator as an ``effective $f_{\rm nl}$,'' its value would be 
\begin{equation}
|f_{\rm nl}| \sim \sqrt{\epsilon} {M_{\rm pl}\over M} . 
\label{fnleff1}
\end{equation}
The sign of this effective $f_{\rm nl}$ depends on the sign of the coefficient $d_1$.  Sometimes it is useful to characterize the two momentum regimes with different effective values of $f_{\rm nl}$---an $f_{\rm nl}^{\rm equil}$ for equilateral momenta and an $f_{\rm nl}^{\rm local}$ for the case of a narrow isosceles triangle.  The current WMAP \cite{wmap} bounds for these two parameters are
\begin{eqnarray}
-151 < &\!\!\!f_{\rm nl}^{\rm equil}\!\!\!& < 253
\nonumber \\
-9 < &\!\!\!f_{\rm nl}^{\rm local}\!\!\!& < 111
\label{wmapfnl}
\end{eqnarray}
where both ranges are quoted at a 95\% confidence level.

As was mentioned, for a fairly general set of symmetry-breaking operators (excluding for the moment those with comparatively many powers of the spatial derivative which we shall discuss next) the three-point function already sets a rather stringent bound on the scale of new physics associated with the operator $H^2\varphi^3$.  For example, even with a rather small value for $\epsilon$, such as $\epsilon\sim 0.01$, $M$ cannot be more than three orders of magnitude below the Planck scale, based on existing bounds.  These constraints should improve with the next set of experiments.  Planck \cite{planck} in particular anticipates measuring $f_{\rm nl}$ to the order of ``a few'' \cite{yadav}.

Of course, how the constraints on the irrelevant operators---based on what we can infer about the two-point and three-point functions from observations---translate into constraints on the overall scale $M$ for the symmetry breaking depends on the particular model.  For a general set of symmetry-breaking operators, we would expect that  
\begin{equation}
{H^3\over M}\varphi^2 ,\quad
{H\over M}\varphi\vec\nabla\cdot\vec\nabla\varphi , \quad
{H^2\over M}\varphi^3 , \quad
\label{irrelseg}
\end{equation}
should all appear with similar coefficients, unless there is some specific principle that suppresses the $\varphi^3$ operator and not the other two.  The main lesson here is that while observations of the power spectrum only weakly constrain the scale $M$ in the first two operators \cite{broken}, bounds on the unobserved three-point function already place much stronger constraints on $M$ as it occurs in the last of these operators.

\subsection{Case II:  The operator $\varphi^2\vec\nabla\cdot\vec\nabla\varphi$}

The amplitudes associated with the second operator, which depends on the spatial derivatives rather than time derivatives (in the form of $H$), have more interesting effects,
\begin{eqnarray}
{\cal A}_2^{\rm equil}(k) 
&\!\!\!\approx\!\!\!& 
{d_2\over 3\sqrt{2}} {1\over\epsilon^{3/2}} {H^3\over M_{\rm pl}^3} 
{k\over k_\star} \sin\biggl[ 3 {M\over H} {k\over k_\star} \biggr] 
\nonumber \\ 
{\cal A}_2^{\rm iso}(k) 
&\!\!\!\approx\!\!\!& 
{3d_2\over 10\sqrt{2}} {1\over\epsilon^{3/2}} {H^3\over M_{\rm pl}^3} 
{k_3\over k_\star} \sin\biggl[ 2 {M\over H} {k\over k_\star} \biggr] . 
\label{A2triangle1}
\end{eqnarray}
Again, for the equilateral case we have assumed that $k_1\sim k_2\sim k_3\sim k$, while for the latter, $k_1\sim k_2\sim k \gg k_3$.  Since the amplitude is proportional to the product of the momenta, $k_1k_2k_3$, the case where one of these is much smaller than the other two (the isosceles triangle) will similarly be much smaller than the case when they are all comparable.  Therefore we shall concentrate on the equilateral case here; when observations are applied to it, we obtain much stronger constraints on the possibilities for the scale of  trans-Planckian---or more accurately, ``trans-Hubble'' ($M>H$)---physics.

Unlike the previous case, the usual scale-independent $f_{\rm nl}$ is not an appropriate approximation here, although we could still define an effective $f_{\rm nl}$ as long as we remember that it only applies to as a particular region of momentum space.  For instance, when they are all nearly equal in magnitude, he have
\begin{equation}
|f_{\rm nl}^{\rm equil}(k)| \sim \sqrt{\epsilon} {M_{\rm pl}\over H} 
{k\over k_\star} \sin\biggl[ 3 {M\over H} {k\over k_\star} \biggr] , 
\label{fnleff2}
\end{equation}
again neglecting the coefficient, $d_2$.  Since we are introducing the scale associated with the symmetry-breaking below the Planck scale, the momentum $k_\star$ that encodes the limit of the applicability of the our effective treatment is now associated with a mode whose wavenumber is equal to $M$ at $\eta_0$, rather than $M_{\rm pl}$ as before.  Therefore, the appropriate range for $k$ is the interval
\begin{equation}
{H\over M} < {k\over k_\star} < 1 . 
\label{koverkstarrange}
\end{equation}
The lower limit is set by the condition that the mode must be within the horizon at the initial time and the upper limit corresponds to the ``trans-Planckian''  (actually $M<M_{\rm pl}$) threshold.  As we noted earlier, when $k\sim k_\star$, higher order operators of the general form
\begin{equation}
{1\over a^{2d}M^{2d-1}} \varphi^2 (\vec\nabla\cdot\vec\nabla)^d \varphi
\label{genopd}
\end{equation}
all have a similar contribution to the three-point function and our effective treatment is simply no longer predictive.  Taking a rough average for this correction, the size of this $f_{\rm nl}^{\rm equil}(k)$ lies within the range, 
\begin{equation}
\sqrt{\epsilon} {M_{\rm pl}\over M} < \bigl| f_{\rm nl}^{\rm equil}(k) \bigr| 
< \sqrt{\epsilon} {M_{\rm pl}\over H} . 
\label{fnlrange}
\end{equation}
Note that the lower end imposes the same constraint on $M$ as the last operator $H^2\varphi^3$.  

The upper end, in contrast, does not depend at all on the symmetry-breaking scale $M$, but instead on the much larger ratio $M_{\rm pl}/H \agt 10^5$.  This limit is only reached if the wavenumber corresponding to a particular physical feature of the universe were generated at the $k\sim aM$ threshold.  However, even in the most conservative case, where the horizon-size fluctuation at the initial time $\eta_0$ is just entering the horizon today, we can observe fluctuations in the microwave background radiation to at least two to three orders of magnitude smaller.  Because this correction grows linearly with $k$, the size of $f_{\rm nl}^{\rm equil}(k)$ at these scales would correspondingly be two to three orders of magnitude larger than its size at the largest wavelengths.  Since observationally, we already know that $|f_{\rm nl}^{\rm equiv}| \alt {\cal O}(10^2)$, this bound would essentially push the symmetry-breaking scale into a genuinely Planckian realm, $M\to M_{\rm pl}$.

\section{Conclusions}
\label{conclude}

Most of the attempts to learn whether and how the trans-Planckian properties of nature might influence inflationary predictions have focused on the power spectrum, or two-point function, of the quantum fluctuations of the background.  The characteristic size of these quantum fluctuations is (neglecting the $k$-dependence) 
\begin{equation}
\zeta_k(0)\propto {1\over\sqrt{\epsilon}} {H\over M_{\rm pl}} , 
\label{fluctsize}
\end{equation}
so it is most natural to look first at the least suppressed correlation functions---that is, those with the fewest factors of the field.  The two-point function is the simplest and largest of these and its influence has already been observed.

Beyond this natural smallness of the fluctuations, the typical inflationary model predicts a pattern that is highly Gaussian in its character, so that the three-point function (and any other ``odd-point'' function) is additionally suppressed.  So while it might be inherently more difficult to observe, new signals beyond those of the traditional inflationary picture stand out more starkly in the three-point function than in the power spectrum.

The correct description of the three-point function is in terms of its Fourier transform, ${\cal A}(k_1,k_2,k_3)$,
\begin{eqnarray}
&&\!\!\!\!\!\!\!\!\!\!\!\!\!\!\!
\langle 0(\eta)|\zeta(\eta,\vec x) \zeta(\eta,\vec y) \zeta(\eta,\vec z) |0(\eta)\rangle
\nonumber \\
&=&
\int {d^3\vec k_1\over (2\pi)^3} {d^3\vec k_2\over (2\pi)^3} 
{d^3\vec k_3\over (2\pi)^3}\, 
e^{i\vec k_1\cdot\vec x} e^{i\vec k_2\cdot\vec y} e^{i\vec k_3\cdot\vec z}
\nonumber \\
&&
(2\pi)^3 \delta^3(\vec k_1+\vec k_2+\vec k_3)
{k_1^3+k_2^3+k_3^3\over k_1^3k_2^3k_3^3}
{\cal A}(k_1,k_2,k_3), \qquad
\label{ftdef}
\end{eqnarray}
which we have normalized with a few factors of momentum, $k_i\equiv |\vec k_i|$, extracted.  More frequently and less accurately, the amount of non-Gaussianity in the pattern of the primordial perturbations is sometimes characterized by a parameter $f_{\rm nl}$ which scales, up to order one constants, as 
\begin{equation}
f_{\rm nl} \sim \epsilon^2 {M_{\rm pl}^4\over H^4} {\cal A} . 
\label{fnlasamp}
\end{equation}
The function ${\cal A}(k_1,k_2,k_3)$ depends on the spatial momenta; this property can be crudely captured by defining different $f_{\rm nl}$'s for different representative regions of the momentum space.  The standard treatment of inflation predicts that $f_{\rm nl}$ should be of the same order as the slow-roll parameters, $f_{\rm nl}\sim \epsilon, \delta$ \cite{nongauss,maldacena}.

This article has examined how new dynamics above the inflationary scale $H$, up to the Planck scale $M_{\rm pl}$, could generate larger non-Gaussianities in the pattern of primordial perturbations.  To keep the analysis fairly general, we have applied an effective treatment to study a theory where some of the space-time symmetries are broken at a scale $M$, $H<M\le M_{\rm pl}$.  In the power spectrum, this scenario, with its broken symmetries, reproduces the signals that occur when the quantum state of the fluctuations contains trans-Planckian structures \cite{effstate}.  At the same time it avoids questions about how to renormalize the theory, since its renormalization is completely standard and contains no further dependence on the state \cite{effstate,tp4,emil,por}.  Considering a generic set of irrelevant three-point operators, an operator of the form $H^2\zeta^3$ yields an effective $f_{\rm nl}$ of 
\begin{equation}
|f_{\rm nl}| \sim \sqrt{\epsilon} {M_{\rm pl}\over M} , 
\qquad\qquad\hbox{[Case I]}
\label{case1rev}
\end{equation}
while the operator $\zeta^2 \vec\nabla\cdot\vec\nabla\zeta$ introduces non-Gaussianities which are {\it minimally\/} of the same size, but which also grow at smaller scales, 
\begin{equation}
|f_{\rm nl}| \sim \sqrt{\epsilon} {M_{\rm pl}\over M} {k\over k_{\rm min}} . 
\qquad\hbox{[Case II]}
\label{case2rev}
\end{equation}
Here $k_{\rm min}$ is a horizon-size fluctuation at the time that we start our time-evolution, and $k$ is the common value of the spatial momenta when they are all about the same size.

The current experimental bounds require $f_{\rm nl}$ to be less than ${\cal O}(100)$ or so.  But fairly soon the Planck satellite with its better precision ought to be able to detect any $f_{\rm nl}$ larger than about ${\cal O}(5)$ \cite{yadav}.  Even the current bounds provide quite strong constraints on the scale of the effects discussed here, and one which is generally much more stringent than those obtained by comparing the predicted ``trans-Planckian'' corrections to the power spectrum to experiments.  With the next series of experiments, the possibility of additional symmetry-breaking, beyond the small amount present in the background, would be pushed into the genuinely---and probably unobservable---trans-Planckian realm, $M\agt M_{\rm pl}$, while operators with many spatial derivatives could probably be ruled out altogether.

Although we have avoided choosing a very specific model by applying an effective treatment of the broken symmetries, we have clearly not considered the most general possible scenario.  It would be interesting to examine other approaches to the trans-Planckian problem, especially those which modify the state of the quantum fluctuations in some way \cite{brandenberger,tp0,tp1}, to see whether they also produce non-Gaussianities of a size similar to what we have found here.

So far, the possible influences of trans-Planckian physics on the predictions of inflation have been described only after choosing a specific state or vacuum for the quantum fluctuations.  While very instructive and able to illustrate what sorts of signals are possible through trans-Planckian effects, this approach does not quite resolve the trans-Planckian problem in its most basic form.  There ought to be a deeper and more universal reason, if the inflationary picture is ultimately correct, why nature forgets what happens at such tiny scales despite the extraordinary expansion in an inflationary universe.

\begin{acknowledgments}

\noindent
This work was supported in part by DOE grant No.~DE-FG03-91-ER40682 and by the Niels Bohr International Academy.

\end{acknowledgments}

\end{document}